\documentclass{elsart}
\usepackage{amsfonts}
\usepackage{amssymb}
\usepackage{amsmath}
\usepackage[dvips]{graphicx}

\begin{document}

\begin{frontmatter}

\title{Cryptanalysis of dynamic look-up table based chaotic cryptosystems}

\author{G. \'{A}lvarez\corauthref{corr}},
\author{F. Montoya},
\author{M. Romera},
\author{G. Pastor}

\corauth[corr]{Corresponding author: Email: gonzalo@iec.csic.es}

\address{Instituto de F\'{\i}sica Aplicada, Consejo Superior de
Investigaciones Cient\'{\i}ficas, Serrano 144---28006 Madrid,
Spain}

\begin{abstract}
In recent years many chaotic cryptosystems based on Baptista's
seminal work have been proposed. We analyze the security of two of
the newest and most interesting ones, which use a dynamically
updated look-up table and also work as stream ciphers. We provide
different attack techniques to recover the keystream used by the
algorithms. The knowledge of this keystream provides the attacker
with the same information as the key and thus the security is
broken. We also show that the dependence on the plaintext, and not
on the key, of the look-up table updating mechanism facilitates
cryptanalysis.
\end{abstract}

\begin{keyword}
Chaotic cryptosystems, Ergodicity, Cryptanalysis, Hash algorithm

\PACS 05.45.Ac, 47.20.Ky.
\end{keyword}

\end{frontmatter}

\mathindent 0cm
\sloppy

\section{Introduction}

Since M. S. Baptista proposed in 1998 a new cryptosystem based on
the property of \emph{ergodicity} of chaotic systems \cite{cwc}, a
number of new algorithms based on variations of Baptista's have
been published \cite{amccm,cwcc,afccswdlut,acccahs,accsfgsc}. In
\cite{coaecc} we analyzed the security and cryptographic
robustness of Baptista's seminal algorithm. The first variation of
Wong \cite{amccm} was cryptanalyzed in \cite{kcoaccm}. We present
in this Letter our results after having thoroughly studied Wong's
second and third algorithms \cite{afccswdlut,acccahs}.

The ergodicity property is exploited in these algorithms by using
the logistic map
\begin{equation}\label{eq:logmap}
    y_{n+1}=by_n(1-y_n),
\end{equation}
where $y_n\in [0,1]$ and the parameter $b$ is chosen so that
Eq.~(\ref{eq:logmap}) behaves chaotically. In \cite{afccswdlut},
the most interesting addition consists of using a dynamic table
for looking up the ciphertext and plaintext which is no longer
fixed during the whole encryption and decryption processes.
Instead, it depends on the plaintext, being continuously updated
during the encryption and decryption processes. This makes
cryptanalysis more difficult, but not impossible.

When the $i$th message block is encrypted, the look-up table is
updated dynamically by exchanging the $i$th entry $l_i$ with
another entry $l_j$ . The location of the latter entry, i.e., the
value of $j$, is determined by the current value of $y$ using the
following formula:
\begin{equation}\label{eq:v}
\upsilon=\left\lfloor{\frac{{y-y_{\min}}}{{y_{\max}-y_{\min}}}}\times
N\right\rfloor
\end{equation}
\begin{equation}\label{eq:lookup}
j=i+\upsilon\mod N,
\end{equation}
where $y_{\min}$ and $y_{\max}$ are the end points of the chosen
interval $[y_{\min},y_{\max})$ and $N$ is the total number of
entries in the table \cite{afccswdlut}.

In \cite{acccahs}, the previously described chaotic cryptographic
scheme is generalized by allowing the swapping of multiple pairs
of entries in the look-up table during the encryption of each
input block, and by allowing multiple runs of encryption on the
whole message continuously. Starting from the current entry $i$,
$p$ pairs of entries ($p\geq 1$) are swapped according to the
following rule: $i\leftrightarrow(i+\upsilon\mod N)$,
$(i+\upsilon+1\mod N)\leftrightarrow(i+2\upsilon+1\mod N)$,
$(i+2\upsilon+2\mod N)\leftrightarrow(i+3\upsilon+2\mod N)$,
\dots, $(i+(p-1)\upsilon+p-1\mod N)\leftrightarrow
(i+p\upsilon+p-1\mod N)$. Once the message has been encrypted, the
whole process is repeated again $r$ times, $r\geq 1$. The final
look-up table is the hash of the message \cite{acccahs}.

\section{Classical types of attacks}

When cryptanalyzing a ciphering algorithm, the general assumption
made is that the cryptanalist knows exactly the design and working
of the cryptosystem under study, i.e., he knows everything about
the cryptosystem except the secret key. This is an evident
requirement in today's secure communications networks, usually
referred to as Kerchoff's principle \cite[p. 24]{ctap}. According
to \cite[p. 25]{ctap}, it is possible to differentiate between
different levels of attacks on cryptosystems. They are enumerated
as follows, ordered from the hardest type of attack to the
easiest:

\begin{enumerate}
    \item Ciphertext only: The opponent possesses a string of
    ciphertext.
    \item Known plaintext: The opponent possesses a string of
plaintext, $\mathbf{p}$, and the corresponding ciphertext,
$\mathbf{c}$.
    \item Chosen plaintext: The opponent has obtained temporary access to the
encryption machinery. Hence he can choose a plaintext string,
$\mathbf{p}$, and construct the corresponding ciphertext string,
$\mathbf{c}$.
    \item Chosen ciphertext: The opponent has obtained temporary access to the
decryption machinery. Hence he can choose a ciphertext string,
$\mathbf{c}$, and construct the corresponding plaintext string,
$\mathbf{p}$.
\end{enumerate}

In each of these four attacks, the objective is to determine the
key that was used. The last two attacks, which might seem
unreasonable at first sight, are very common when the
cryptographic algorithm, whose key is fixed by the manufacturer
and unknown to the attacker, is embedded in a device which the
attacker can freely manipulate. Daily life examples of such
devices are smartcards, electronic purse cards, GSM phone SIM
(Subscriber Identity Module) cards, POST (Point Of Sale Terminals)
machines, or web application session token encryption.

\section{Keystream attacks}
\label{sec:pad}

Although at first sight the cipher under study might look like a
block cipher, in fact it behaves as a stream cipher \cite[p.
20]{ctap}, a fundamental weakness as is to be seen. The operation
of the algorithm as a stream cipher can be explained as follows.
Suppose $K$ is the key, given by $y_0$ and $b$, and that
$\mathbf{p}=p_1p_2\ldots$ is the plaintext string. A keystream
$\mathbf{k}=k_1k_2\ldots$ is generated using
Eq.~(\ref{eq:logmap}). This keystream is used to encrypt the
plaintext string according to the rule
\[\mathbf{c}=e_{k_1}(p_1)e_{k_2}(p_2)\ldots=c_1c_2\ldots\]
Decrypting the ciphertext string $\mathbf{c}$ can be accomplished
by computing the keystream $\mathbf{k}$ given the knowledge of the
key $K$ and undoing the operations $e_{k_i}$. In \cite{afccswdlut}
the keystream is the complete orbit followed by iterating
Eq.~(\ref{eq:logmap}) from the initial point $y_0$ with parameter
near $b=4.0$. The unit interval is divided up into $N$ equally
spaced bins, each corresponding to a symbol of the alphabet in
use. However, instead of considering the whole unit interval, only
the subinterval $[0.2,0.8)$ is used. As a consequence of the
natural invariant density of Eq.~(\ref{eq:logmap}), the orbit will
visit frequently the forbidden subintervals $[0,0.2)$ and
$[0.8,1]$.

As an example of how to generate the keystream, let us iterate
Eq.~(\ref{eq:logmap}) starting from $y_0=0.1777$ and
$b=3.9999995$, as in \cite{afccswdlut}. Let $s_i$ be the symbol
corresponding to each useful subinterval where $y_i$ lands, and
let $x$ be the iterates which visit the forbidden subintervals.
The orbit followed is $y_i=$\{0.5844\dots, 0.9714\dots,
0.1109\dots, 0.3945\dots, 0.9555\dots, 0.1698\dots, 0.5641\dots,
0.9835\dots, 0.0647\dots, 0.2420\dots, \ldots\}, and is
transformed into the keystream $\mathbf{k} =
s_{165}xxs_{84}xxs_{156}xxs_{18}\ldots$ Next we show how to
recover the keystream using chosen ciphertext, chosen plaintext,
and known plaintext attacks. It is important to note that knowing
the keystream $\mathbf{k}$ generated by a certain key $K$ ($y_0$
and $b$) is entirely equivalent to knowing the key. Therefore, our
keystream attacks focus on recovering $\mathbf{k}$.

\subsection{How to circumvent the look-up table}

The election of the look-up table updating method is most
unfortunate, since it allows the attacker to easily predict the
new positions of the symbols even without the exact knowledge of
the value of $y$. In order to initially simplify our analysis, we
use a variable number $N$ of symbols, $N=2^n,n=1,\ldots,8$. First,
we assume that the source emits two different symbols, $S_2=\{
s_1,s_2\}$. When the orbit lands on the first subinterval
$[0.2,0.5)$, the table will be updated following Eqs.~(\ref{eq:v})
and (\ref{eq:lookup}). Given that $0.2\leq y<0.5$, we have that
\[
0\leq\upsilon=\left\lfloor\frac{{y-y_{\min}}}{{y_{\max}-y_{\min}}}\times
2\right\rfloor=\left\lfloor\frac{{y-0.2}}{{0.6}}\times
2\right\rfloor<1,
\]
and thus $\upsilon=0$ and Eq.~(\ref{eq:lookup}) is equivalent to:
\[j=i\mod 2.\]
When the orbit lands on the second subinterval $[0.5,0.8)$, then
$0.5\leq y<0.8$, and thus $\upsilon=1$ and Eq.~(\ref{eq:lookup})
becomes:
\[j=(i+1)\mod 2.\]
If the source emits four different symbols,
$S_4=\{s_1,s_2,s_3,s_4\}$, then $\upsilon=0$ if the orbit lands on
[0.2,0.35), $\upsilon=1$ if the orbit lands on [0.35,0.5),
$\upsilon=2$ if the orbit lands on [0.5,0.65), and $\upsilon=3$ if
the orbit lands on [0.65,0.8).

The generalization for higher order sources is immediate. Even
when multiple pairs are swapped at each encryption run, as in
\cite{acccahs}, the look-up table evolution is easily predicted.
As a consequence, the look-up table plays no significant security
role during the encryption process. It is not necessary to know
the exact value of $y$ to predict the next update. It suffices to
know the subinterval where $y$ lands. Therefore, the updated table
depends solely on the plaintext, and not on the key ($y_0$ and
$b$), to the advantage of the attacker. When encrypting the same
plaintext using different keys, the same updating sequence will
take place for the look-up table. Likewise, the same message will
always yield the same hash value regardless of the key used.

\subsection{How to obtain the keystream}

In this subsection different attacks are described. Each of them
aims at the recovery of the keystream.

\subsubsection{Chosen ciphertext attack}

The chosen ciphertext attack is straightforward. Simply request
the plaintext of the one-block ciphertexts $\mathbf{c}=1$,
$\mathbf{c}=2$, $\mathbf{c}=3$, \ldots, until the desired length
of the keystream is reached. Either the correct symbol (when the
iterate lands on a site) or an error (when it lands outside the
boundaries) is obtained, one by one. Once the desired length of
keystream has been recovered in this way, any message encrypted
with the same values of $y_0$ and $b$ can easily be decrypted.
Under this attack, the dynamically updated look-up table has no
effect at all. This attack requires as many one-block ciphertexts
as the length of the keystream that is to be recovered. This
attack does not work on \cite{acccahs}, because the attacker does
not know the encrypted values of $p$ and $r$, the first two blocks
of the ciphertext.

\subsubsection{Chosen plaintext attack}

Let us deal with the chosen plain text attack next. To make things
even simpler at first, we assume the fourth order symbol source
$S_4$ and that $r_{\max}=1$ (see \cite{afccswdlut}). Although
unknown to the cryptanalyst, the system key $K$ is given by
$y_0=0.1777$ and $b=3.9999995$, using the interval $[0.2,0.8)$, as
in \cite{afccswdlut}.

Our goal is to find out the exact position of all occurrences of
$s_1$,$s_2$,$s_3$, and $s_4$ in the keystream. But this task is
not as easy as requesting the ciphertext of
$\mathbf{p}=s_1s_1s_1\dots$, then of $\mathbf{p}=s_2s_2s_2\dots$,
etc., as in \cite{coaecc}. The dynamic look-up table prevents
knowing whether a certain symbol corresponds to its original
position in the table or has been already changed. But given that
the changes produced in the table are known even when the exact
value of $y$ is unknown, the following attack can be designed.

First, in order to know the exact position of all symbols $s_1$ in
the keystream $\mathbf{k}$, we need to construct an adequate
plaintext $\mathbf{p}$. Table~1 represents the final result of the
process followed to compute the correct value of $\mathbf{p}$. We
start constructing Table~1 by filling in columns $i$ and $k_i$,
already known in advance. Columns 0, 1, 2, and 3 reflect the
current state of the look-up table, i.e., which symbol is at which
position at any given moment. Next, we proceed row by row, in the
following way:

\begin{enumerate}
    \item Assign to $p_i$ the value of the symbol ($s_1$, $s_2$, $s_3$, or
    $s_4$) in the previous row in the subinterval corresponding to $k_i$. At start, the look-up table is
    not yet altered.
    \item Calculate $j=(i+k_i)\mod 4$.
    \item Update the look-up table by interchanging the symbols in the subintervals $i,j$.
    \item Proceed to the next row.
\end{enumerate}

After proceeding in this way, the plaintext that corresponds to
all symbols $s_1$ in the keystream is worked out:
$\mathbf{p}=0\;0\;0\;0\;0\dots$ This plaintext is always periodic.
The corresponding ciphertext is requested, obtaining:
$\mathbf{c}=10\;9\;6\;6\;7\dots$. Hence, it is known for sure that
there is a true symbol $s_1$ at the 10th position, and at the
19th, etc. The partial keystream already obtained is
$\mathbf{k}=xxxxxxxxxs_1xxxxxxxxs_1xxxxxs_1xxxxxs_1xxxxxxs_1\dots$

Next, we are to obtain the exact position of all symbols $s_2$ by
constructing Table~2. This table informs us that we have to
request the ciphertext of
$\mathbf{p}=1\;0\;2\;2\;2\;0\;3\;3\;3\;0\;1\;1\dots$, obtaining
$\mathbf{c}=4\;11\;14\;5\dots$ The improved partial keystream
already obtained is
$\mathbf{k}=xxxs_2xxxxxs_1xxxxs_2xxxs_1xxxxxs_1xxxs_2xs_1xxs_2xxxs_1\dots$

If we repeat this process, generating Table~3 and 4 for symbols
$s_3$ and $s_4$, and requesting the corresponding ciphertexts, we
would obtain the following complete keystream:
\begin{equation}\label{eq:keystream}
\mathbf{k}=s_3xxs_2xxs_3xxs_1s_4s_4s_4xs_2xxxs_1s_4s_4xs_3xs_1s_4s_4xs_2xs_1s_4xs_2xxxs_1\dots
\end{equation}
The generalization for higher order sources is immediate. This
attack is very inexpensive too, since it only requires $N=2^n$
plaintexts, $1\leq n\leq 8$.

When $r_{\max}>1$, the same procedure must be followed. However,
there will be blanks in the recovered keystream, because many
valid iterations will be skipped. In many cases, though, it is
possible to narrow down the number of possible symbols. Once the
partial keystream has been worked out, while trying to decrypt a
ciphertext following the decryption method, iterates will land on
an $x$. The only possible symbols for those $x$ are those lying
before the $x$ at a distance smaller than $r_{\max}$. In this way,
the possibilities are greatly reduced, in many cases to only one
possible value (the correct one). When more than one value is
possible, there are two courses of action. If the plaintext is not
of random nature, then the gaps can easily be filled selecting by
the context one symbol amongst the possible ones. On the other
hand, if the plaintext is random, then a new plaintext must be
requested, made by all the previous correct symbols plus one of
the candidates. If the ciphertext is equal to the expected one,
then the guess was correct. Otherwise, new plaintexts must be
encrypted until the correct guess is used.

There is still another possible modification presented in
\cite{afccswdlut}. A new parameter, called threshold
($y_{\mathrm{threshold}}$), can be introduced along with the key.
The current value of $y$ can be checked against this threshold, so
that the table is updated only when $y>y_{\mathrm{threshold}}$.
However, although it is assumed that this addition improves
security, it is very easy to deduce which is the subinterval to
which $y_{\mathrm{threshold}}$ belongs. It can be observed in
Table~1 that when the plaintext symbol to be encrypted is $s_1$,
the table is never effectively updated. After the $N=2^n$ tables
are constructed, if $y_{\mathrm{threshold}}>y_{\min}$, then there
must occur repeated values in the recovered keystream. Given that
the position of $s_1$ is always correct in the keystream, we know
for sure that symbols $s_t$ for $t>1$ which coincide with a
previous $s_1$ symbol must be incorrect. The greatest value of $t$
indicates which is the symbol $y_{\mathrm{threshold}}$ belongs to.
This attack works on \cite{acccahs} too. The attacker can still
predict the evolution of the look-up table with the only knowledge
of the plaintext.

\subsubsection{Known plaintext attack}

Under this attack, each plaintext/ciphertext pair allows for the
recovery of a portion of the keystream. For the sake of simplicity
and without loss of generality, let us use once again the $S_4$
source, $r_{\max}=1$, and $y_{\mathrm{threshold}}=y_{\min}$. Let
us set $\mathbf{p}=1\;0\;3\;0\;0\dots$, whose ciphertext is
$\mathbf{c}=4\;11\;5\;9\;5\dots$ Table~5 can be easily
constructed, which allows to recover a correct portion of the
keystream:
$\mathbf{k}=xxxs_2xxxxxxxxxxs_2xxxxs_4xxxxxxxxs_2xxxxs_2\dots$
This process should be repeated with as many known plaintexts as
possible, to recover as big a portion of the keystream as
possible. Therefore, this attack does not guarantee total recovery
of the keystream. It would work in a similar way for
\cite{acccahs}.

\subsection{How to decrypt using the keystream}

In the previous subsection, different methods to obtain the
complete keystream where introduced. Next, we explain how to
recover the plaintext from a ciphertext when the keystream is
known.

For simplicity, the fourth order symbol source $S_4$ is used. We
assume the attacker already knows the keystream given by
Eq.~(\ref{eq:keystream}) and possesses the following ciphertext:
$\mathbf{c}=1\;10\;4\;4\;4\dots$ In order to decrypt it, Table~6
is constructed in the following way. First, fill all the values of
$k_i$ with the symbol found in the keystream at the positions
indicated by the ciphertext. Next, according to the current state
of the look-up table, assign the correct value to $p_i$ which
corresponds to each $k_i$. Calculate the value of $j$ and update
the look-up table accordingly. Move to the next row and repeat the
process until the ciphertext has been exhausted.

\section{Security of the hash}

We have tested that breaking the hash algorithm is possible when
$p=1$ and $r=1$, even without the knowledge of the key ($y_0$ and
$b$). Let us consider for example the following message:
``Transfer \$10000 to Alvarez's account.''. If encoded using 4-bit
symbols, its hash is $h=\mathrm{1E825BC0A36974FD}$, expressed in
hexadecimal. Changing the message into ``Transfer \$30005 to
Alvarez's account.'' produces exactly the same hash. In order to
avoid attacks on the hashing scheme, it is all important that
$r>1$ and $p>1$. In effect, in \cite{acccahs} two runs and a small
value of $p$ are suggested. Greater values would increase
security, but penalize on speed.

Although in \cite{acccahs} it is claimed that this hash can be
treated as a message authentication code (MAC), in fact it can
not. A MAC is a key-dependent one-way hash function. However, as
already proved, the look-up table, and hence the hash, does not
depend on the key ($y_0$ and $b$). Therefore, this scheme does not
behave as a MAC but as a one-way hash function, even though the
knowledge of the key is necessary to verify the hash when both
authenticity and secrecy are to be provided.

\section{Conclusions}

In spite of dynamically updating the look-up table, the same
fundamental weakness present in Baptista's algorithm \cite{cwc} is
reproduced in the chaotic cryptosystems proposed in
\cite{afccswdlut,acccahs}, as proved by our different attacks. As
a consequence of our attacks, an important conclusion is that
implementations of these algorithms can never reuse the same key
because if so, they are easily broken. Furthermore, the look-up
table does not depend on the key, but only on the plaintext, thus
facilitating cryptanalysis. After these attacks, we conclude that
the lack of security, along with the low encryption speed,
discourage the use of these algorithms for secure applications. We
are to investigate how the weaknesses outlined in this Letter
might affect the security of other Wong's variants
\cite{accsfgsc}.

\ack{This research was supported by Ministerio de Ciencia y
Tecnolog\'{\i}a, Proyecto TIC2001-0586. Our thanks to K. W. Wong
for providing his source code.}

\clearpage

\begin{table}
  \centering
  \caption{Plaintext ($p_i$) needed to find out the exact position of symbols $s_1$ ($k_i=s_1=0$).}
  \begin{tabular}{rrrrrrrr}
  \hline
  $i$ & $j$ & $0$ & $1$ & $2$ & $3$ & $k_i$ & $p_i$ \\
  \hline
-  & - & $s_1$ & $s_2$ & $s_3$ & $s_4$ & - &  -   \\
0 & 0 & $s_1$ & $s_2$ & $s_3$ & $s_4$ & 0 & $s_1$ \\
1 & 1 & $s_1$ & $s_2$ & $s_3$ & $s_4$ & 0 & $s_1$ \\
2 & 2 & $s_1$ & $s_2$ & $s_3$ & $s_4$ & 0 & $s_1$ \\
3 & 3 & $s_1$ & $s_2$ & $s_3$ & $s_4$ & 0 & $s_1$ \\
  \hline
  \end{tabular}
\end{table}

\begin{table}
  \centering
  \caption{Plaintext ($p_i$) needed to find out the exact position of symbols $s_2$ ($k_i=s_2=1$).}
  \begin{tabular}{rrrrrrrr}
  \hline
  $i$ & $j$ & $0$ & $1$ & $2$ & $3$ & $k_i$ & $p_i$ \\
  \hline
-  & - & $s_1$ & $s_2$ & $s_3$ & $s_4$ & - &  -   \\
0 & 1 & $s_2$ & $s_1$ & $s_3$ & $s_4$ & 1 & $s_2$ \\
1 & 2 & $s_2$ & $s_3$ & $s_1$ & $s_4$ & 1 & $s_1$ \\
2 & 3 & $s_2$ & $s_3$ & $s_4$ & $s_1$ & 1 & $s_3$ \\
3 & 0 & $s_1$ & $s_3$ & $s_4$ & $s_2$ & 1 & $s_3$ \\
$4\equiv 0$ & 1 & $s_3$ & $s_1$ & $s_4$ & $s_2$ & 1 & $s_3$ \\
$5\equiv 1$ & 2 & $s_3$ & $s_4$ & $s_1$ & $s_2$ & 1 & $s_1$ \\
$6\equiv 2$ & 3 & $s_3$ & $s_4$ & $s_2$ & $s_1$ & 1 & $s_4$ \\
$7\equiv 3$ & 0 & $s_1$ & $s_4$ & $s_2$ & $s_3$ & 1 & $s_4$ \\
$8\equiv 0$ & 1 & $s_4$ & $s_1$ & $s_2$ & $s_3$ & 1 & $s_4$ \\
$9\equiv 1$ & 2 & $s_4$ & $s_2$ & $s_1$ & $s_3$ & 1 & $s_1$ \\
$10\equiv 2$ & 3 & $s_4$ & $s_2$ & $s_3$ & $s_1$ & 1 & $s_2$ \\
$11\equiv 3$ & 0 & $s_1$ & $s_2$ & $s_3$ & $s_4$ & 1 & $s_2$ \\
  \hline
  \end{tabular}
\end{table}

\begin{table}
  \centering
  \caption{Plaintext ($p_i$) needed to find out the exact position of symbols $s_3$ ($k_i=s_3=2$).}
  \begin{tabular}{rrrrrrrr}
  \hline
  $i$ & $j$ & $0$ & $1$ & $2$ & $3$ & $k_i$ & $p_i$ \\
  \hline
-  & - & $s_1$ & $s_2$ & $s_3$ & $s_4$ & - &  -   \\
0 & 2 &  $s_3$ & $s_2$ & $s_1$ & $s_4$ & 2 & $s_3$ \\
1 & 3 &  $s_3$ & $s_4$ & $s_1$ & $s_2$ & 2 & $s_1$ \\
2 & 0 & $s_1$ & $s_4$ & $s_3$ & $s_2$ & 2 & $s_1$ \\
3 & 1 & $s_1$ & $s_2$ & $s_3$ & $s_4$ & 2 & $s_3$ \\
  \hline
  \end{tabular}
\end{table}

\begin{table}
  \centering
  \caption{Plaintext ($p_i$) needed to find out the exact position of symbols $s_4$ ($k_i=s_4=3$).}
  \begin{tabular}{rrrrrrrr}
  \hline
  $i$ & $j$ & $0$ & $1$ & $2$ & $3$ & $k_i$ & $p_i$ \\
  \hline
-  & - & $s_1$ & $s_2$ & $s_3$ & $s_4$ & - &  -   \\
0 & 3 &  $s_4$ & $s_2$ & $s_3$ & $s_1$ & 3 & $s_4$ \\
1 & 0 &  $s_2$ & $s_4$ & $s_3$ & $s_1$ & 3 & $s_1$ \\
2 & 1 & $s_2$ & $s_3$ & $s_4$ & $s_1$ & 3 & $s_1$ \\
3 & 2 & $s_2$ & $s_3$ & $s_1$ & $s_4$ & 3 & $s_1$ \\
$4\equiv 0$ & 3 & $s_4$ & $s_3$ & $s_1$ & $s_2$ & 3 & $s_4$ \\
$5\equiv 1$ & 0 & $s_3$ & $s_4$ & $s_1$ & $s_2$ & 3 & $s_2$ \\
$6\equiv 2$ & 1 & $s_3$ & $s_1$ & $s_4$ & $s_2$ & 3 & $s_2$ \\
$7\equiv 3$ & 2 & $s_3$ & $s_1$ & $s_2$ & $s_4$ & 3 & $s_2$ \\
$8\equiv 0$ & 3 & $s_4$ & $s_1$ & $s_2$ & $s_3$ & 3 & $s_4$ \\
$9\equiv 1$ & 0 & $s_1$ & $s_4$ & $s_2$ & $s_3$ & 3 & $s_3$ \\
$10\equiv 2$ & 1 & $s_1$ & $s_2$ & $s_4$ & $s_3$ & 3 & $s_3$ \\
$11\equiv 3$ & 2 & $s_1$ & $s_2$ & $s_3$ & $s_4$ & 3 & $s_3$ \\
  \hline
  \end{tabular}
\end{table}

\begin{table}
  \centering
  \caption{Partial keystream ($k_i$) recovered when both plaintext ($p_i$) and ciphertext ($c_i$) are known.}
  \begin{tabular}{rrrrrrlrr}
  \hline
  $i$ & $j$ & $0$ & $1$ & $2$ & $3$ & $k_i$ & $p_i$ & $c_i$ \\
  \hline
-  & - & $s_1$ & $s_2$ & $s_3$ & $s_4$ & - &  - & -  \\
0 & 1 &  $s_2$ & $s_1$ & $s_3$ & $s_4$ & $s_2=1$ & $s_2$ & 4 \\
1 & 2 &  $s_2$ & $s_3$ & $s_1$ & $s_4$ & $s_2=1$ & $s_1$ & 11 \\
2 & 1 & $s_2$ & $s_1$ & $s_3$ & $s_4$ & $s_4=3$ & $s_4$ & 5 \\
3 & 0 & $s_4$ & $s_1$ & $s_3$ & $s_2$ & $s_2=1$ & $s_1$ & 9 \\
$4\equiv 0$ & 1 & $s_1$ & $s_4$ & $s_3$ & $s_2$ & $s_2=1$ & $s_1$ & 5 \\
  \hline
  \end{tabular}
\end{table}

\begin{table}
  \centering
  \caption{Plaintext ($p_i$) recovered when the keystream ($k_i$) is known.}
  \begin{tabular}{rrrrrrlr}
  \hline
  $i$ & $j$ & $0$ & $1$ & $2$ & $3$ & $k_i$ & $p_i$ \\
  \hline
-  & - & $s_1$ & $s_2$ & $s_3$ & $s_4$ & - &  -   \\
0 & 2 & $s_3$ & $s_2$ & $s_1$ & $s_4$ & $s_3=2$ & $s_3$ \\
1 & 0 & $s_2$ & $s_3$ & $s_1$ & $s_4$ & $s_4=3$ & $s_4$ \\
2 & 3 & $s_2$ & $s_3$ & $s_4$ & $s_1$ & $s_2=1$ & $s_3$ \\
3 & 3 & $s_2$ & $s_3$ & $s_4$ & $s_1$ & $s_1=0$ & $s_2$ \\
$4\equiv 0$ & 2 & $s_4$ & $s_3$ & $s_2$ & $s_1$ & $s_3=2$ & $s_4$ \\
  \hline
  \end{tabular}
\end{table}


\begin{thebibliography}{1}

\bibitem{cwc}
M.~S. Baptista.
\newblock Cryptography with chaos.
\newblock {\em Physics Letters A}, 240:50--54, 1998.

\bibitem{amccm}
W. Wong, L. Lee, and K. Wong.
\newblock A modified chaotic cryptographic method.
\newblock {\em Computer Physics Communications}, 138:234--236, 2001.

\bibitem{cwcc}
A.~Palacios and H.~Juarez.
\newblock Crytography with cycling chaos.
\newblock {\em Physics Letters A}, 303:345--351, 2002.

\bibitem{afccswdlut}
K. W. Wong.
\newblock A fast chaotic cryptographic scheme with dynamic look-up table.
\newblock {\em Physics Letters A}, 298:238--242, 2002.

\bibitem{acccahs}
K. W. Wong.
\newblock A combined chaotic cryptographic and hashing scheme.
\newblock {\em Physics Letters A}, 307:292--298, 2003.

\bibitem{accsfgsc}
K. W. Wong, S. W. Ho, and C. K. Yung.
\newblock A chaotic cryptography scheme for generating short ciphertext.
\newblock {\em Physics Letters A}, 310:67--73, 2003.

\bibitem{coaecc}
G. \'{A}lvarez, F. Montoya, M. Romera, and G. Pastor.
\newblock Cryptanalysis of an ergodic chaotic cipher.
\newblock {\em Physics Letters A}, 311:172--179, 2003.

\bibitem{kcoaccm}
G. \'{A}lvarez, F. Montoya, M. Romera, and G. Pastor.
\newblock Keystream cryptanalysis of a chaotic cryptographic method.
\newblock {\em Computer Physics Communications}, in press.

\bibitem{ctap}
D.~R. Stinson.
\newblock {\em Cryptography: theory and practice}.
\newblock CRC Press, 1995.

\end{thebibliography}
\end{document}